\documentstyle[pra,aps]{revtex}

\begin{document}
\title{Gravitational quantum limit for length measurements }
\author{Marc-Thierry Jaekel $^a$ and Serge Reynaud $^b$}
\address{$(a)$ Laboratoire de Physique Th\'eorique de l'ENS \thanks{%
Unit\'e propre du CNRS associ\'ee \`a l'Ecole Normale Sup\'erieure et \`a %
l'Universit\'e Paris-Sud}, 24 rue Lhomond F75231 Paris Cedex 05 France\\
$(b)$ Laboratoire de Spectroscopie Hertzienne de l'ENS \thanks{%
Unit\'e de l'Ecole Normale Sup\'erieure et de l'Universit\'e Pierre et Marie
Curie associ\'ee du CNRS }, 4 place Jussieu F75252 Paris Cedex 05 France}
\date{{\sc Physics Letters} {\bf A 185} (1994) 143-148}
\maketitle

\begin{abstract}
We discuss a limit for sensitivity of length measurements which is due to
the effect of vacuum fluctuations of gravitational field. This limit is
associated with irreducible quantum fluctuations of geodesic distances and
it is characterized by a noise spectrum with an order of magnitude mainly
determined by Planck length. The gravitational vacuum fluctuations may (in
an analysis restricted to questions of principle and when the measurement
strategy is optimized) dominate fluctuations added by the measurement
apparatus if macroscopic masses, i.e. masses larger than Planck mass, are
used.
\end{abstract}

\section{Introduction}

Quantum fluctuations put limits on the sensitivity in length measurements.
For measurements based upon electromagnetic probes, sensitivity appears at
first sight to be bound by the 'standard quantum limit', a compromise
between length fluctuations on one hand and momentum fluctuations of
end-points on the other hand \cite{bib1}. Stated in other words, phase
fluctuations of the electromagnetic probe appear as fluctuations of the
measured length, while intensity fluctuations give rise to a random motion
of the end-points (for instance mirrors of an interferometer), through the
action of radiation pressure. Since phase and intensity are quantum
conjugated variables with fluctuations bound by an Heisenberg inequality, a
standard quantum limit is derived when considering their contributions as
uncorrelated \cite{bib2}.

However, phase and intensity fluctuations are not independent noise sources,
and it is feasible, in principle, to push sensitivity beyond standard
quantum limit \cite{bib3}. It turns out that sensitivity is limited only by
dissipation in the mechanical admittance of end-points \cite{bib4}.
Focussing the attention upon questions of principle and putting discussion
of experimental problems aside (thermal fluctuations, limitations for
optical or mechanical finesses, control of fluctuations of the
electromagnetic probe, ..), one is then led to consider fundamental
dissipation mechanisms which cannot be bypassed. One such mechanism, the
radiation pressure of vacuum fluctuations, leads to a sensitivity limit \cite
{bib5} which is of the order of the Compton wavelength $\lambda _{C}$
associated with the end-points' masses.

This result could be expected from a mere dimensional analysis using the
Planck constant $\hbar $, the light velocity $c$ and the mass $m$. As soon
as gravity is considered however, Planck units of mass and length may be
defined ($G$ is the Newton's gravitational constant):
\begin{eqnarray}
m_{p} &=&\sqrt{\frac{\hbar c}{G}} \approx 2.2\ 10^{-8}\text{kg}
  \nonumber \\
l_{P} &=&ct_{p}=\frac{c}{\omega _{p}}=\sqrt{\frac{\hbar G}{c^{3}}}
\approx 1.6\ 10^{-35}\text{m}  \label{eq1}
\end{eqnarray}
and sensitivity in length measurements is expected to be limited by Planck
length (see for example \cite{bib6}). This is in conflict with the analysis
sketched above, since Planck length is larger than Compton wavelength for
masses greater than Planck mass. The purpose of the present letter is to
show that a limit of the order of Planck length indeed appears when quantum
fluctuations of gravitational field are taken into account.

The phase of an electromagnetic probe registers curvature perturbations
associated with classical gravitational waves. In particular, it can detect
a stochastic background of gravitational waves (\cite{bib7} and references
in) which may have been generated by various astrophysical or cosmological
mechanisms (\cite{bib8} and references in). Here, we will consider only the
effect of gravitational vacuum fluctuations, i.e. proper quantum
fluctuations of gravitational field. These fluctuations may be derived (for
frequencies smaller than Planck frequency) in a linearized theory of
gravitation in the same way as vacuum fluctuations of other physical fields
(see for instance \cite{bib9}). We show in the following that they limit the
sensitivity of geodesic measurements to Planck length. It has to be noted in
particular that, though small, these fluctuations dominate vacuum pressure
fluctuations for masses greater than Planck mass. This change in dominant
fluctuations supports the intuition that Planck mass might be considered as
a natural borderline between microscopic and macroscopic domains.

\section{Quantum limits in a length measurement}

For completeness, we summarize in the present section the main results known
for quantum limits in length measurements performed with an electromagnetic
probe. The standard quantum limit is characterized as an ultimate detectable
length variation $\Delta q$:
\begin{equation}
\left( \Delta q^{2}\right) ^{SQL}\approx \frac \hbar m T
\label{eq2}
\end{equation}
where $T$ is the measurement time. It can be derived by considering that the
endpoints' positions are measured at two times separated by an interval $T$
and noting that positions at different times do not commute \cite{bib10}.
This derivation is implicitly based upon too restrictive assumptions and it
consequently does not lead to the ultimate limit \cite{bib11}.

To present a more detailed analysis of ultimate limits, it is appropriate to
describe length fluctuations by a time correlation or a noise spectrum
defined according to the prescription:
\begin{eqnarray}
C_{qq}\left( t\right) &\equiv &\left\langle q\left( t\right) q\left(
0\right) \right\rangle -\left\langle q\left( t\right) \right\rangle
\left\langle q\left( 0\right) \right\rangle  \nonumber \\
&\equiv &\int
{\displaystyle {d\omega  \over 2\pi }}%
C_{qq}\left[ \omega \right] e^{-i\omega t}  \label{eq3}
\end{eqnarray}
Ascribing the noise to phase and intensity fluctuations of the
electromagnetic probe \cite{bib2}, one obtains the standard quantum limit
when considering their contributions as statistically uncorrelated. Assuming
that the signal is monitored at frequencies where the end-points are nearly
free, one thus gets the noise spectrum \cite{bib4}:
\[
C_{qq}^{SQL}\left[ \omega \right] \approx \frac{\hbar }{m\omega ^{2}}
\]
This corresponds to a noise energy per unit bandwidth of the order of $\hbar
$. The variance $\Delta q^{2}$ may then be evaluated for a measurement
bandwidth $\Delta \omega $:
\begin{equation}
\left( \Delta q^{2}\right) ^{SQL}\approx \frac{\hbar }{m}\frac{\Delta \omega
}{2\pi \omega ^{2}}  \label{eq4}
\end{equation}
The expression (\ref{eq2}) is recovered when $\frac{\Delta \omega }{2\pi
\omega ^{2}}$ is interpreted as the detection time $T$.

Now, the contributions of phase and intensity fluctuations are linearly
superimposed in the monitored signal, and the total noise can be reduced
\cite{bib3} by squeezing the appropriate field quadrature component
(references on squeezing may be found in \cite{bib12}). Assuming that the
field fluctuations around the probe frequency can be squeezed at
convenience, one demonstrates that sensitivity is limited only by the
dissipative part of the mirrors' mechanical admittance \cite{bib4}. The same
analysis shows that standard quantum limit is determined by the reactive
part, and is therefore revealing the inadequacies of a measurement strategy
rather than ultimate quantum limitations. This conclusion is consistent with
a general analysis of the effects of noise and dissipation in
high-sensitivity measurements \cite{bib13}.

A lower bound for dissipation is set by radiation pressure of vacuum fields,
characterized by a spectrum $C_{FF}$ for the force $F$ exerted upon the
mirrors:
\[
C_{FF}\left[ \omega \right] =\frac{\hbar ^{2}}{3\pi c^{2}}\omega ^{3}\theta
\left( \omega \right) \Phi \left[ \omega \right]
\]
The dimensionless function $\Phi $ is equal to $1$ for a perfect mirror in a
model world with only one spatial dimension \cite{bib14}. In
four-dimensional space-time, diffraction effects are present and \`{e}
depends on geometric characteristics of the mirror \cite{bib15}. For a
perfectly reflecting mirror with a characteristic dimension $\rho $, $\Phi $
is found to be greater than $1$ for $\omega \rho \gg c$ and smaller than $1$
for $\omega \rho \ll c$, the latter case corresponding to a more realistic
situation ($\omega $ is a mechanical frequency). Note that the function $%
\Phi $be enhanced by resonance effects in a cavity \cite{bib16}.

As a result of these force fluctuations, the mirrors undergo a random motion
characterized by a noise spectrum \cite{bib5}:
\[
C_{qq}\left[ \omega \right] =\frac{C_{FF}\left[ \omega \right] }{m^{2}\omega
^{4}}
\]

This random motion determines a 'vacuum-pressure quantum limit' which lies
far beyond standard quantum limit (\ref{eq4}), with an order of magnitude
given by the Compton wavelength $\lambda _{C}$:
\begin{equation}
C_{qq}^{VQL}\left[ \omega \right] =\frac{\Phi }{3\pi }\lambda _{C}^{2}\frac{%
\theta \left( \omega \right) }{\omega }\qquad \lambda _{C}=\frac{\hbar }{mc}
\label{eq5}
\end{equation}
Except for the factor $\frac{\Phi }{3\pi }$, these expressions may be
infered from simple dimensional arguments. They represent the ultimate
sensitivity in a length measurement which would only be limited by quantum
fluctuations of a measurement apparatus build with masses $m$.

\section{Quantum fluctuations of space-time curvature}

As discussed in the introduction, the phaseshift of the electromagnetic
probe between the two end-points registers modifications of the space-time
curvature associated with gravitational fluctuations, including in
particular quantum fluctuations. For gravitational wave detectors, such
fluctuations constitute the quantum fluctuations of the monitored signal
itself.

Gravity fluctuations are usually represented as fluctuations of the metric $%
h_{\mu \nu }$ written in a particular gauge \cite{bib7,bib8,bib9}. In order
to emphasize the irreducible character of these space-time fluctuations, we
prefer here to write manifestly Lorentz-invariant and gauge-independent
correlation functions for linearized Riemann curvature fluctuations $R_{\mu
\nu \rho \sigma }$ (more details on their derivation from the graviton
propagator may be found in \cite{bib17}; $k$ is a 4-wavevector, and $k_{0}$
the corresponding frequency; the spectrum $C[k]$ is the 4-dimensional
Fourier transform of the space-time correlation function $C(x)$, defined as
in eq.(\ref{eq3}); $\eta _{\mu \nu }$ is the Minkowski metric tensor with a
signature (+,-,-,-)):
\begin{eqnarray}
C_{R_{\mu \nu \rho \sigma }R_{\mu ^{\prime }\nu ^{\prime }\rho ^{\prime
}\sigma ^{\prime }}}\left[ k\right]  &=&16\pi ^{2}l_{P}^{2}\theta \left(
k_{0}\right) \delta \left( k^{2}\right)   \nonumber \\
&\times &\left\{ {\cal R}_{\mu \nu \mu ^{\prime }\nu ^{\prime }}{\cal R}%
_{\rho \sigma \rho ^{\prime }\sigma ^{\prime }}+{\cal R}_{\mu \nu \rho
^{\prime }\sigma ^{\prime }}{\cal R}_{\rho \sigma \mu ^{\prime }\nu ^{\prime
}}-{\cal R}_{\mu \nu \rho \sigma }{\cal R}_{\mu ^{\prime }\nu ^{\prime }\rho
^{\prime }\sigma ^{\prime }}\right\}   \nonumber \\
{\cal R}_{\mu \nu \rho \sigma } &=&\frac{1}{2}\left( k_{\mu }k_{\rho }\eta
_{\nu \sigma }+k_{\nu }k_{\sigma }\eta _{\mu \rho }-k_{\nu }k_{\rho }\eta
_{\mu \sigma }-k_{\mu }k_{\sigma }\eta _{\nu \rho }\right)   \label{eq6}
\end{eqnarray}
As classical gravitational waves, gravitational vacuum fluctuations are
concentrated upon the light cone (lightlike wavevectors) and correspond to
vanishing Einstein curvature. Note that equations (\ref{eq6}) give only the
lowest order contribution proportional to $l_{P}^{2}$ (equivalently to $G$)
to curvature fluctuations. Higher order contributions exist, due in
particular to gravity of vacuum stress-tensor fluctuations. Their evaluation
would in principle require a fully consistent theory of Quantum Gravity.
They have however a much smaller magnitude when evaluated at frequencies
much lower than Planck frequency $\omega _{P}$ (they scale at least as $%
l_{P}^{4}$), so that we shall ignore them here.

For simplicity, equations (\ref{eq6}) and subsequent computations are
written with natural space-time units ($c=1$) and light velocity is
reintroduced in the final expressions.

\section{Quantum fluctuations of geodesic distances}

As freely falling end-points follow geodesics, we will evaluate variations
of the geodesic distance (measured as an electromagnetic phaseshift) by
using the law of geodesic deviation \cite{bib18}. Such a law gives the
'equivalent tidal acceleration' between points on two neighbouring geodesics
as the product of their distance by the curvature component $R_{0\mu 0\nu
}u^\mu u^\nu $ where $u$ is the electromagnetic wavevector (normalized in
such a manner that $u_0 = 1$). In the general case of a finite distance
between the two end-points, the geodesic deviation is obtained as an
integral along the path of the probe.

For a measurement on a 'one-way track', the variation q of geodesic distance
due to curvature fluctuations is obtained from ($x$ is the 4-coordinate of
the receiver, $t=x_{0}$ the corresponding time, $\tau $ the time of
propagation from the emitter to the receiver, $\sigma $ the affine parameter
along the path):
\begin{equation}
\frac{d^{2}q}{dt^{2}}=\int_{0}^{\tau }R_{0\mu 0\nu }\left( x-u\sigma \right)
u^{\mu }u^{\nu }d\sigma  \label{eq7}
\end{equation}

This relation is often written with $q(t)$ expressed in terms of metric
components (usually in the transverse-traceless gauge) or with the redshift
parameter $\left( -\frac{dq}{dt}\right) $ expressed in terms of time
derivatives of such metric components (see \cite{bib7} and references in;
also compare with the Sachs-Wolfe formula \cite{bib19}).

After a translation into the frequency domain, one obtains the noise
spectrum $C_{qq}$ describing the fluctuations of geodesic distance as an
integral over gravitational wavevectors. Using the expressions (\ref{eq6})
describing the gravitational vacuum fluctuations, one then derives a
'gravitational quantum limit':
\begin{equation}
C_{qq}^{GQL}\left[ \omega \right] =l_{P}^{2}\frac{\theta \left( \omega
\right) }{\omega }b\left[ \omega \right]  \label{eq8}
\end{equation}
where the dimensionless function $b$ (denoted $b^{\left( 1\right) }$ for
one-way tracking) is:
\[
b^{\left( 1\right) }\left[ \omega \right] =\left\langle \left( 1+\gamma
\right) ^{2}\left| e^{i\omega \tau }-e^{i\gamma \omega \tau }\right|
^{2}\right\rangle
\]
Here, $\gamma $ is the cosine of the spatial angle between gravitational
wavevector $k$ and electromagnetic propagation direction $u$ ($\gamma =\frac{%
k_{m}u^{m}}{k_{0}}$; note that $\omega =k_{0}$) and the symbol $<>$
represents an angular average (i.e. the average over $\gamma $ uniformly
distributed between $-1$ and $1$) so that:
\[
b^{\left( 1\right) }\left[ \omega \right] =\frac{8}{3}-\frac{4}{\left(
\omega \tau \right) ^{2}}+\frac{2\sin \left( 2\omega \tau \right) }{\left(
\omega \tau \right) ^{3}}
\]
In the case of 'two-way tracking', the distance is measured as half a round
trip time, which leads to a different expression for the function $b$:
\begin{eqnarray*}
b^{\left( 2\right) }\left[ \omega \right] &=&\frac{1}{4}\left\langle \left|
\left( 1+\gamma \right) \left( e^{i\omega \tau }-e^{i\gamma \omega \tau
}\right) -\left( 1-\gamma \right) \left( e^{-i\omega \tau }-e^{i\gamma
\omega \tau }\right) \right| ^{2}\right\rangle \\
&=&1-\frac{\cos \left( 2\omega \tau \right) }{3}-\frac{3+\cos \left( 2\omega
\tau \right) }{\left( \omega \tau \right) ^{2}}+\frac{2\sin \left( 2\omega
\tau \right) }{\left( \omega \tau \right) ^{3}}
\end{eqnarray*}

At the low-frequency limit (propagation length smaller than gravitational
wavelength), the same expression is obtained in both cases of one-way and
two-way tracking (a local tidal acceleration is measured in this case):
\begin{eqnarray*}
&&b^{\left( 1\right) }\left[ \omega \right] \approx b^{\left( 2\right)
}\left[ \omega \right] \approx \left( \omega \tau \right) ^{2}\left\langle
\left( 1-\gamma ^{2}\right) ^{2}\right\rangle \qquad \left( \omega \tau \ll
1\right) \\
&&C_{qq}^{GQL}\left[ \omega \right] \approx \frac{8}{15}l_{P}^{2}\tau
^{2}\omega \theta \left( \omega \right)
\end{eqnarray*}
At this limit, geodesic fluctuations may be reproduced by fluctuations of a
local length scale, i.e. by conformal metric fluctuations. However, this
picture is no longer valid for measurements involving several propagation
directions. For example, in the geometry of a Michelson interferometer with
two propagation directions, correlation functions depend on the angle
between the two propagation directions (see \cite{bib8} and references in).

At the high-frequency limit in contrast (propagation length larger than
gravitational wavelength), fluctuations are found to be independent of the
propagation time \cite{bib20}:
\begin{eqnarray*}
C_{qq}^{GQL}\left[ \omega \right] &\approx &l_{P}^{2} \frac{\theta
\left( \omega \right)} \omega
b\left[ \infty \right] \qquad \left( \omega \tau \gg
1\right) \\
b^{\left( 1\right) }\left[ \infty \right] &=&\frac{8}{3}\qquad b^{\left(
2\right) }\left[ \infty \right] =1
\end{eqnarray*}
At this limit, fluctuations of geodesic distances may be reproduced by
independent stochastic motions of the two end-point positions.

\section{Commutation relations for geodesic distance}

It has to be noted that the functions $b$ appearing in equation (\ref{eq8})
are obtained in the same manner for any isotropic spectrum of gravitational
fluctuations (see \cite{bib7} and references in) while the other factors
appearing in expression (\ref{eq8}) are characteristic of vacuum
fluctuations. In particular, equations (\ref{eq8}) describe fluctuations of
non-commuting variables, as the curvature fluctuations (\ref{eq6}) from
which they are deduced (through the law of geodesic deviation (\ref{eq7})).
This is made more visible by writing the commutator:
\[
\left[ q\left( t\right) ,q\left( 0\right) \right] =C_{qq}\left( t\right)
-C_{qq}\left( -t\right)
\]
or, in the spectral domain:
\[
C_{qq}\left[ \omega \right] -C_{qq}\left[ -\omega \right] =l_{P}^{2}\frac{%
b\left[ \omega \right] }{\omega }
\]
($b$ is an even function). Note that this commutator, computed here in the
vacuum state (of gravity waves), is state-independent. The same value would
be obtained in a stochastic background of gravity waves (at the level of
approximation considered in this letter).

The commutator between the distance variation $q$ and the associated
velocity $q^{\prime }$ (which is also the redshift parameter) is thus
directly related to the function $b$:
\[
C_{q^{\prime }q}\left[ \omega \right] -C_{qq^{\prime }}\left[ -\omega
\right] =il_{P}^{2}b\left[ \omega \right]
\]
Translating back to the time domain provides us with commutators describing
quantum fluctuations of geodesic distances:
\begin{eqnarray}
\left[ q^{\prime }\left( t\right) ,q\left( 0\right) \right]
&=&-il_{P}^{2}b\left( t\right)  \nonumber \\
\left[ q\left( t\right) ,q\left( 0\right) \right] &=&-il_{P}^{2}B\left(
t\right)  \nonumber \\
B\left( t\right) &=&\int_{0}^{t}b\left( t^{\prime }\right) dt^{\prime }
\label{eq9}
\end{eqnarray}
Explicit expressions of these functions are obtained for one-way tracking:
\begin{eqnarray*}
b^{\left( 1\right) }\left( t\right) &=&\frac{8}{3}\delta \left( t\right) -%
\frac{\left( 2\tau -\left| t\right| \right) ^{2}}{2\tau ^{3}}\theta \left(
2\tau -\left| t\right| \right) \\
B^{\left( 1\right) }\left( t\right) &=&\varepsilon \left( t\right) \frac{%
\left( 2\tau -\left| t\right| \right) ^{3}}{6\tau ^{3}}\theta \left( 2\tau
-\left| t\right| \right)
\end{eqnarray*}
and for two-way tracking:
\begin{eqnarray*}
b^{\left( 2\right) }\left( t\right) &=&\delta \left( t\right) -\frac{1}{6}%
\left( \delta \left( t-2\tau \right) +\delta \left( t+2\tau \right) \right) +%
\frac{\left( \left| t\right| -\tau \right) \left( 2\tau -\left| t\right|
\right) }{2\tau ^{3}}\theta \left( 2\tau -\left| t\right| \right) \\
B^{\left( 2\right) }\left( t\right) &=&\varepsilon \left( t\right) \frac{%
-2\left| t\right| ^{3}+9\left| t\right| ^{2}\tau -12\left| t\right| \tau
^{2}+6\tau ^{3}}{12\tau ^{3}}\theta \left( 2\tau -\left| t\right| \right)
\end{eqnarray*}
The short-time dynamics is independent of the mean propagation time (as the
high-frequency limit). A remarkable property of the commutator is that it
contains a step at null delay whose size is given by the high-frequency
limit $b\left[ \infty \right] $ of the function $b$:
\[
B\left( t\right) \approx \frac{1}{2}b\left[ \infty \right] \varepsilon
\left( t\right) \qquad \left( \left| t\right| \ll \tau \right)
\]

\section{Discussion}

Quantum fluctuations of gravitational fields set a limit for sensitivity in
length measurements. This limit corresponds to irreducible fluctuations for
geodesic distances, which have a quantum character even in empty space. It
follows that a consistent treatment of vacuum fluctuations and gravitation
must take into account the non-commutative quantum geometry of space-time
\cite{bib21}.

The quantitative expressions (\ref{eq8}) (or (\ref{eq9})) obtained for the
gravitational quantum limit have an order of magnitude mainly determined by
Planck length. This is consistent with the often-expressed idea (see for
instance \cite{bib6}) that Planck length limits the range of validity of a
classical description of space-time. Let us however emphasize that we have
characterized the space-time fluctuations for noise frequencies much smaller
than Planck frequency. As a consequence, we can study their (small) effects
in the domain of experimentally accessible frequencies, although we do not
yet have at our disposal a complete theory of Quantum Gravity (which would
describe their large effects at the Planck frequency). In particular, we can
consider the possibility of (small) modifications in the effective
(low-frequency) theory of gravitation, or of quantum fields.

An interesting conclusion about the significance of Planck mass emerges from
a comparison between gravitational quantum limit (\ref{eq8}) and
vacuum-pressure quantum limit (\ref{eq5}). They have rather similar forms if
we consider as unimportant the explicit forms of the dimensionless functions
$\Phi$ and $b$. The main difference between them is that their orders of
magnitude are determined respectively by Planck length $l_P$ and Compton
wavelength $\lambda_C$. It follows that vacuum-pressure fluctuations
dominate for masses smaller than Planck mass while gravitational quantum
fluctuations dominate for masses greater than Planck mass.

At this point, it is worth emphasizing that Planck mass does not appear such
an unaccessible value as Planck length or Planck frequency (see eqs (\ref
{eq1})). Remarkably enough, Planck mass lies near the borderline between
microscopic and macroscopic masses. It is thus tempting to regard this
property, not as an accidental coincidence, but as a consequence for masses
greater than Planck mass of the dominance of a universal fluctuation
mechanism associated with gravitational vacuum fluctuations \cite{bib22}.

Concerning the theory of measurement, it appears that an apparatus designed
for length measurement is able (in an analysis restricted to questions of
principle and if the measurement strategy is optimized) to register the
proper quantum fluctuations of geodesic distances only if it is built with
masses greater than Planck mass. Otherwise, it adds fluctuations of its own
which are greater than geodesic fluctuations by a ratio $\frac{m_{P}}{m}$.
It may be stated in other words that, when quantum fluctuations (and not
only classical trajectories) are considered, macroscopic masses (but not
microscopic ones) are found to obey the principle of universality of free
fall.

\end{document}